\documentclass[pre,showpacs,twocolumn,superscriptaddress,epsfig,longeq]{revtex4}

\usepackage{color}
\usepackage{graphicx}
\usepackage{color}
\graphicspath{{./figures/}}

\begin{document}

\title{Universality and Scaling Behaviour of Injected Power in Elastic Turbulence in Worm-like Micellar Gel}
\author{Sayantan Majumdar} 
\affiliation{Department of Physics, Indian Institute of Science,Bangalore 560012, INDIA}    
\author{A.K. Sood}
\email[]{asood@physics.iisc.ernet.in}                                           
\affiliation{Department of Physics, Indian Institute of Science,Bangalore 560012, INDIA}             
\date{\today}
\begin{abstract}
We study the statistical properties of spatially averaged global injected power fluctuations for Taylor-Couette flow of a worm-like micellar gel formed by surfactant CTAT. At sufficiently high Weissenberg numbers (Wi) the shear rate and hence the injected power $p(t)$ at a constant applied stress shows large irregular fluctuations in time. The nature of the  probability distribution function (pdf) of $p(t)$ and the power-law decay of its power spectrum are very similar to that observed in recent studies of elastic turbulence for polymer solutions. Remarkably, these non-Gaussian pdfs can be well described by an universal large deviation functional form given by the Generalized Gumbel (GG) distribution observed in the context of spatially averaged global measures in diverse classes of highly correlated systems. We show by in-situ rheology and polarized light scattering experiments that in the elastic turbulent regime the flow is spatially smooth but random in time, in agreement with a recent hypothesis for elastic turbulence.  

\end{abstract}
\maketitle

Hydrodynamic instability like turbulence is an inertia driven phenomenon occurring at high 
Reynolds numbers (Re). However, for viscoelastic fluids of long polymers, similar instabilities 
have been observed even at a very low Re because of elastic hoop stresses generated by the stretching of the polymers in the curvilinear flow field, indicated by high Weissenberg number (Wi) \cite{VSNature}. Wi defined by the ratio of time scale set by the relaxation time $(\tau_R)$ of the fluid to that set by the strain rate ($\stackrel{.}{\gamma}$),  Wi = $\tau_R\stackrel{.}{\gamma}= N_{1}/\sigma$, with $N_{1}$ is the first normal stress difference at a shear stress $\sigma$, plays the same role for elastic turbulence that Re plays in the case of inertial turbulence. The elastic turbulence \cite{VSNature} is a flow instability that occurs at practically zero Re (Re$<<$1) and high Wi (Wi$>>$1).

In past few years the statistics and scaling properties of the injected power fluctuations have been studied in detail both experimentally and numerically for inertial turbulence \cite{Fauve, Abry, Cadot, Labbe, Pinton, Titon} in von Karman swirling flows. It has been shown that the probability distribution function (pdf) of spatially averaged global injected power shows deviation from the Gaussian nature, skewness being on the side of smaller value of the injected power. Inertial turbulence also shows spatially smooth but random in time flow behaviour like elastic turbulence, below the Kolmogorov dissipation scale. Recently, similar results have emerged for power fluctuations in polymers in the context of elastic turbulence in von Kerman flow \cite{VSPRL09}. Although, the sign of the asymmetry of the power fluctuations appears to be same for both the inertial and the elastic turbulence, the magnitude of the normalized third moment or skewness vary in different studies. For some cases like in \cite{Pinton} the pdfs are very skewed (skewness $\sim$ -1) but in many other cases 
\cite{Titon, VSPRL09} this value is $\sim$ -0.2. It has been argued that averaging over many independent large scale structures decreases the skewness of the global quantities. Very recently, the flow behaviour of worm-like micellar system has been shown to have signatures of elastic turbulence \cite{SLElasticTurbulence}. To our knowledge, there is no similar study of statistics and universality properties of spatially averaged global injected power in the case of unstable flow at high Wi and very low Re for worm-like micellar gels. These studies are important for worm-like micellar systems, since worm-like micelles have additional relaxation mechanisms due to scission-reattachment dynamics compared to polymers \cite{Cates}.    
This motivated us to study the spatially averaged global injected power fluctuations for Taylor-Couette flow of a shear thinning worm-like micellar solution which shows elastic turbulence. At sufficiently high Wi, the time-series of injected power $p(t) = \sigma\stackrel{.}{\gamma}(t)V_s$ (where $\sigma$ is the applied stress, $\stackrel{.}{\gamma}(t)$ is the resulting shear rate and $V_s$ is the volume of the sample) at a fixed shear stress shows large irregular fluctuations. The pdfs of $p(t)$ deviate from the Gaussian nature with skewness toward low power side with respect to $\left\langle p(t)\right\rangle$ and the values of skewness  obtained are similar to that in Newtonian fluids \cite{Titon} and polymer flows \cite{VSPRL09} despite the differences in the flow profiles in von Karman and Taylor-Couette geometries. Moreover, the three systems discussed above, namely, a Newtonian fluid \cite{Titon}, polymer solutions \cite{VSPRL09} and the present worm-like micellar gel follow different routes to turbulence from the laminar flow because of their completely different dynamics and relaxation mechanisms which make their statistical similarities in the turbulent regime even more non-trivial. We show for the first time, that the non-Gaussian pdf of the injected power $p(t)$ for elastic turbulence in worm-like micellar system can be described by an universal large deviation functional form given by the Generalized-Gumbel (GG) distribution, observed for fluctuations of injected power in inertial turbulent flows \cite{BramwellNature, Pinton, Portelli} and many other correlated systems \cite{Joubaud, electro1, electro2, 1/f, tokamak, thinfilm, Brey, glass,imbibition}. Physically, when the correlation length becomes comparable to the system size, the system can not be divided into a large number of independent domains for any global or spatially averaged measurements, hence, central limit theorem no longer holds and deviations from the Gaussian statistics is observed. To see the existence of this correlation directly, we performed in-situ polarized light scattering and rheology experiments as described in detail in \cite{SayantanPRE}. In the turbulent regime, the scattered light signals obtained from the gap of the transparent Couette cell show random fluctuations. A calculation of spatial and temporal correlation functions of the scattered intensity fluctuations reveals that although the time correlation is very short ranged, the system has very long ranged spatial correlation, practically limited by the size of the Couette cell itself. By this direct method we show the non-trivial space-time flow behaviour satisfying the criterion for the elastic turbulence in micellar system. 
To test the statistical robustness of the results, we repeated all the measurement in another Couette geometry with completely different aspect ratio. Both sets of results are in very good agreement, with each other.          

We used CTAT 2 wt\% + 100 mM NaCl in water which forms a worm-like micellar gel for our experiments. The preparation of the sample is discussed in detail in \cite{RajeshPRL}. All the experiments were carried out on a MCR300 stress controlled rheometer (Anton PAAR, Germany) at a temperature of 26 $^{0}$C. For imaging experiments, we used home made transparent Couette geometry having inner cylinder diameter of 23 mm and height 40 mm and with a gap of 1 mm (Couette A). The outer cylinder is made of transparent glass and are partially enclosed by a water circulation chamber to control the temperature (the exposed portions of the glass cell were used for imaging). A thin sheet of polarized laser beam, formed by placing a cylindrical lens and polarizer arrangement in front of a randomly polarized He-Ne laser (10 mW) was sent along the vorticity $(\nabla \times \textbf{v})$ direction to illuminate the entire gap in $(\nabla \times \textbf{v}, \nabla v)$ plane and imaging was done perpendicularly to the illuminated plane with a wide angle lens and a CCD camera (Lumenera 0.75C, 640 $\times$ 480 pixels). By this arrangement the polarization of the incident laser beam can be tuned continuously in $(\textbf{v}, \nabla v)$ plane. All the optical components were mounted on XYZ stages for precision controls. For good statistical accuracy, we analyze the shear-rate/stress time series with almost 50000 data points (data points during the start-up transients were ignored) sampled at a frequency of 5 Hz. The experimental tolerence in the value of applied shear-stress is $\sim 10^{-3}$\% and for applied shear-rate (obtained through a feed-back mechanism in our stress controlled instrument) is $\sim$1\%.  

\begin{figure}[htbp]
\includegraphics[width=8.9 cm]{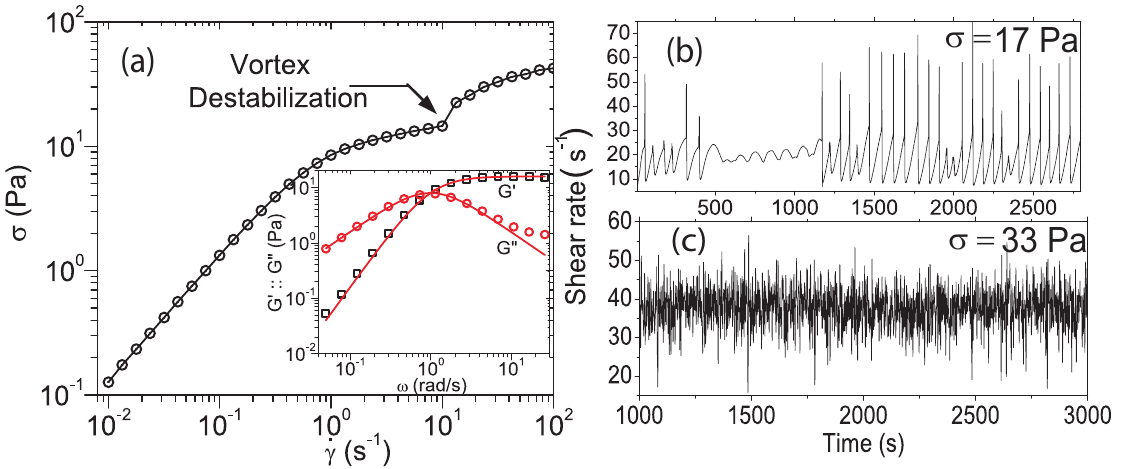}
\caption{(a) Shear stress ($\sigma$) as a function of shear rate ($\stackrel{.}{\gamma}$) for CTAT 2 wt\% + 100 mM NaCl. The inset shows the Storage (G') and loss (G") modulii as a function of angular frequency $\omega$, the solid lines indicate the fit to the single relaxation Maxwell's model. Shear-rate as a funtion of time for different apllied shear stresses (b) $\sigma$ = 17 Pa, (c) $\sigma$ = 33 Pa.} 
\label{Figure1}
\end{figure}

The flow curve for CTAT 2wt\% + 100mM NaCl is shown in Fig.1a. The system shows Maxwellian behaviour \cite{Cates} with a single relaxation time $\tau_R\sim$ 1 s (inset to Fig.1a). In the non-linear regime, in situ imaging experiments do not show shear banding, but reveal Taylor vortex flow right from the onset of the shear-thinning plateau. Also, at $\stackrel{.}{\gamma}\sim$ 10, there is a sudden jump in the stress value, indicating the enhanced flow resistance because of the onset of elastic turbulence. Beyond this value of $\stackrel{.}{\gamma}$ the Taylor vortices get destabilized and move randomly manifesting in large fluctuations in shear rate at a constant applied shear stress and vice versa \cite{SayantanPRE, SayantanEPJB}. We study the temporal dynamics of the system in the non-linear regime for different applied shear stress values of 17 Pa and 33 Pa. The shear-rate does not exhibit fluctuations at stress values less than 17 Pa. For stress value of 17 Pa which corresponds to onset of elastic turbulence, shear-rate shows intermittent behaviour indicating that the system becomes unstable and shows huge fluctuations as shown in Fig.1b. Fig.1c indicates that the shear-rate fluctuates randomly with time for a stress value of 33 Pa. This is the elastic turbulence regime and all the further statistical analysis are done in this regime. The dynamics of shear-rate exactly correlates with the dynamics of the Taylor vortices as observed in \cite{SayantanPRE}. 

\begin{figure*}[htbp]
\includegraphics[width=16 cm]{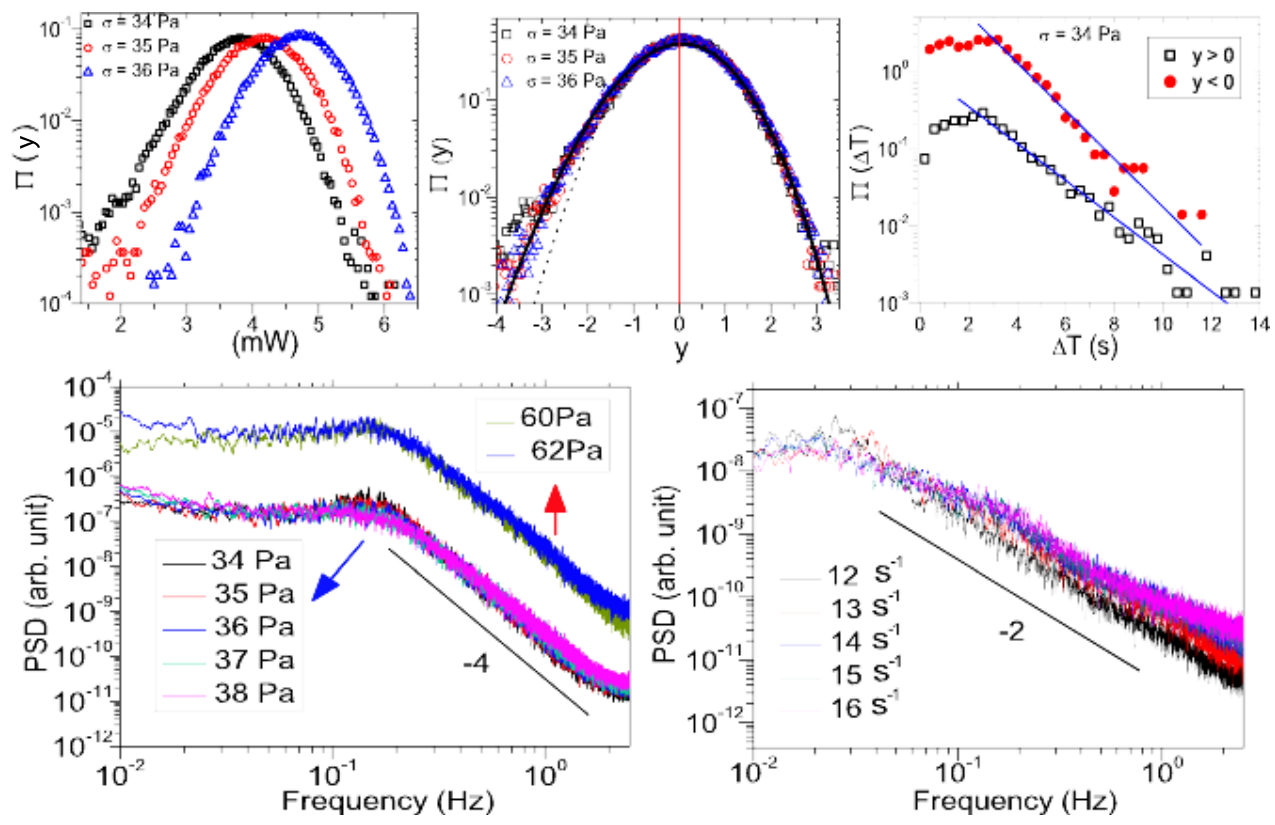}
\caption{Pdfs $\Pi(p)$ and $\Pi(y)$ of injected power for different applied stress values in the elastic turbulent regime, $\sigma$ = 34 Pa, 35 Pa and 36 Pa in terms of (a) raw and (b) reduced (defined in the text) injected power. The solid line in (b) indicates the simulated curve using GG distribution for a = 39 and the dotted line is the fit to Gaussian function. (c) Pdf $\Pi(y)$ of spike width for $y>$0 (hollow squares) and $y<$0 (solid circles) for an applied stress ($\sigma$) of 34 Pa. The solid lines indicate the exponential fits ($\Pi(\Delta T)=\Pi_{0} e^{-\Delta T/\tau}$) to the data. For clarity pdf for $y<$0 have been shifted by a multiplicative factor of 10. 
(d) PSD of injected power fluctuations for applied stress values of 32 Pa, 33 Pa, 34 Pa, 35 Pa and 36 Pa (for Couette A) and also for 60 and 62 Pa (for Couette B, described in the text) The solid straight line indicates a power-law exponent of -4. The curves for Couette B are shifted by a multiplicative factor of 100 for clarity. (e) PSD of injected power in constant shear-rate mode for 12$s^{-1}$, 13$s^{-1}$, 14$s^{-1}$, 15$s^{-1}$ and 16$s^{-1}$ (Couette A). The solid straight line indicates a power-law exponent of -2.} 
\label{Figure2}
\end{figure*} 
 
The pdfs $\Pi(p)$ of injected power ($p$) in the turbulent regime for different applied shear stress values $\sigma$ = 34 Pa, 35 Pa and 36 Pa are shown in Fig.2a. From Fig.2a we see that with increasing $\sigma$, the peak of the corresponding pdf shifts toward higher value of the injected power. In Fig.2b the same pdfs are plotted in terms of reduced value of injected power $y = \frac{\Delta P}{\Gamma}$, where, $\Delta P = p(t) - \left\langle p(t)\right\rangle$. Here, $\left\langle p(t)\right\rangle$ and $\Gamma$ are respectively the mean and the standard deviation of $p(t)$. It can be seen that all the curves collapse into a master curve which deviate from the Gaussian function with negative skewness. The dotted line is a fit to Gaussian function to show the non-Gaussian nature of the master pdf. We now show that the non-Gaussian pdf of injected power $p(t)$ shown in Fig.2b is well represented by Generalized Gumbel distribution (GG) given by \cite{BramwellPRL, CluselEPL}, 

\begin{equation}
\Pi(y) =  K(a) e^{a[-b(a)(y+z(a))-e^{-b(y+z(a))}]}
\end{equation}

\begin{figure*}[htbp]
\includegraphics[width=17 cm]{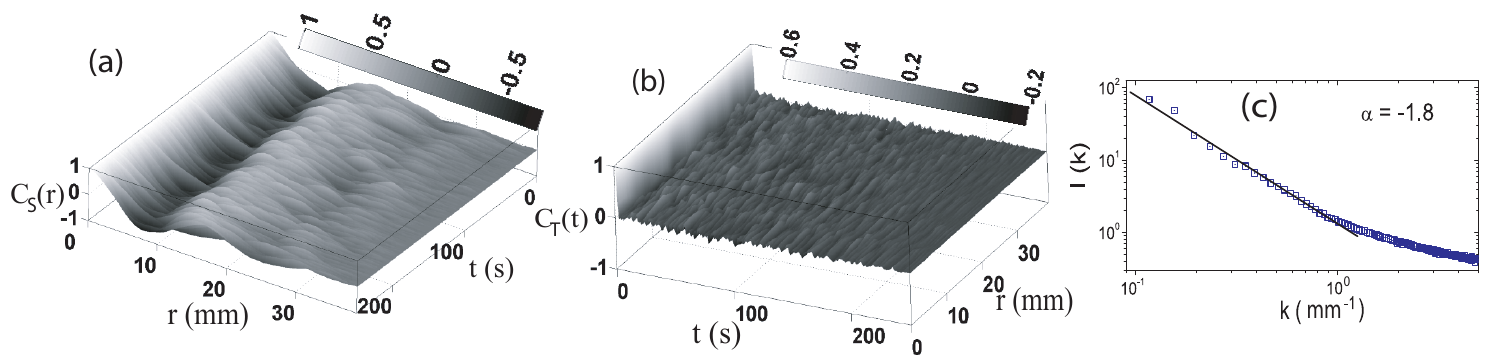}
\caption{Space-time behaviour of (a) normalized spatial correlation function $C_{S}(r)$ and (b) normalized temporal correlation function $C_{T}(t)$. (c) Time averaged spatial Fourier-amplitude I(k) of scattered intensity as a function of wave vector k (=1/l), the solid line is a power-law fit with exponent -1.8 to the data. The applied stress $\sigma = 33$ Pa in all cases.} 
\label{Figure3}
\end{figure*}

The above equation has a single parameter '$a$' which is again fixed by the skewness of the distribution as $\left\langle y^3\right\rangle = 1/\sqrt{a}$. All other parameters are related to '$a$' as follows, ${b(a)}^2 = \frac{d^2ln\Gamma(a)}{da^2}$, $z(a) = \frac{1}{b}(lna - \frac{dln\Gamma(a)}{da})$ and $K(a) = \frac{ba^a}{\Gamma(a)}$ where $\Gamma(a)$ is the Gamma function. The non-Gaussian pdf with negative skewness matches perfectly to the simulated curve given by the GG distribution (Eq.1) as shown by the solid lines in Fig.2b over $\sim$ 3 orders of magnitude with the value of the parameter a = 39, estimated from the third moment of the experimentally measured reduced injected power $\left\langle y^3\right\rangle$. Fig.2c shows the distribution of temporal spike width ($\Delta T$) (persistence time of a single fluctuation event) for the reduced injected power y above and below its mean value ($\left\langle y\right\rangle$ = 0) for an applied stress of 34 Pa. For both the cases $\Pi(\Delta T)$ shows exponential tails over two orders of magnitude as shown by the linear fits in log-linear scale, indicating the statistical independence of the large time events, although for short times they remain strongly correlated. The time constants for the exponential decays are $\tau$ = 2.2 s for $y>0$ and 1.4 s for $y<0$. The mean spike width $\left\langle\Delta T\right\rangle$ = 2.8 s for $y>0$ and 2.6 s for $y<0$. Since, the skewness of $\Pi(y)<0$ (Fig.2b), the higher values of $\left\langle\Delta T\right\rangle$ and $\tau$ for $y>0$ compared to those for $y<0$ implies that the large intermittent negative fluctuations have shorter life-time.The skewness of the distribution on the negative side has also been seen in polymer solutions \cite{Burghelea} which has been postulated to be due to accumulation of the elastic stress near the static boundary. This observation is counterintuitive to the fact that that the shear-rate is higher near the inner cylinder and hence would result in larger stresses near the inner cylinder. More work is needed to understand our results.

The power spectral density (PSD) of $p(t)$ (i.e. $\stackrel{.}{\gamma}(t)$) at different applied stress values in the turbulent regime are shown in Fig.2d, clearly displaying a power-law decay with an exponent $\beta$ = -4 over five orders of magnitude and a smeared peak at $\sim$0.15 Hz. The value of $\beta$ satisfy the Fouxon \cite{Fouxon} criterion ($\left|\beta\right|>$3) for the elastic turbulence. To see the statistical robustness of the exponent $\beta$, we repeated all the experiments in another Couette geometry having different aspect ratio: inner cylinder diameter of 32 mm and height 16.5 mm and a gap of 2 mm (Couette B). Here, the onset of elastic turbulence occurs at much higher value of applied stress ($\sim$ 50 Pa). The difference in the onset of elastic turbulence can come from the difference in curvatures of the two geometries.  The value of $\beta$ obtained for the PSD evaluated at the applied stress values of 60 and 62 Pa is again -4, as shown in Fig.2d. The PSD of $p(t)$ (i.e. $\sigma (t)$) at constant shear-rate for Couette A is shown in Fig.2e where again a power law with a slope of -2 is observed which does not satisfy the Fouxon criterion. We do not understand the origin of different power laws seen in controlled shear stress and controlled shear rate experiments.

To look for any long range correlations in the system in the turbulent regime, we did in-situ polarized light scattering studies as described in detail in \cite{SayantanPRE}. We image the gap of the Couette cell A as a function of time in $(\nabla \times \textbf{v}, \nabla v)$ plane in the turbulent regime with a CCD camera at a frame  rate of 1 Hz. This frame rate is sufficient to capture the correlated dynamics in the system, since the injected power $p(t)$ has a correlation time of $\sim$ 3 s. The image of the entire gap ($\sim40\times$1 mm), is represented by a slice of 370x11 pixels and we analyze 250 such slices in the turbulent regime. We study the space-time distribution of scattered intensity $I(r,t)$, by studying the intensity along a vertical cut which covers the entire length of the cell along the vorticity direction (370x1 pixels) chosen near the center of the gap of the Couette cell and then stacking them at different times as the Space-Time Plot (STP). The STP of the scattered intensity distribution in the turbulent regime is quite random (data not shown). The normalized spatial correlation functions $C_{S}(r) = \left\langle I(r')I(r + r')\right\rangle_{r'}$ evaluated at different times in the turbulent regime are shown in Fig.3a which clearly reveals that the scattered intensity is correlated spatially over the entire length of the Couette geometry and practically limited by the size of the container, at every t. Similarly, we also calculate the normalized temporal correlation function $C_{T}(t) = \left\langle I(t')I(t + t')\right\rangle_{t'}$ for all the pixels along the cut. The correlation function $C_{T}(t)$ is extremely short lived, almost delta-correlated (up to our time resolution) as shown in Fig.3b. The temporal correlation function monotonously goes to the base-line over $\sim$ 1 s as has been confirmed by a higher image grabbing rate of 3 Hz (not shown). These observations clearly indicate that although the spatially averaged injected power fluctuates randomly in time, the spatial structures remain intact and they move randomly in time. This spatially smooth and temporarily random flow behaviour supports a recent hypothesis of elastic turbulence, not reported so far, in the context of micellar gels. Fig.3c shows the plot of time averaged spatial Fourier-transform of scattered intensity along the vertical cut mentioned above which show a power law decay (I(k) $\sim$ k$^{\alpha}$) with exponent $\alpha = -1.8$ (similar to that obtained in \cite{SLElasticTurbulence}) over a length scale spanning $\sim$ 1 to 10 mm.  
     
In conclusion, we have studied the universality and scaling properties of spatially averaged global injected power $p(t)$ in the context of elastic turbulence in a shear-thinning worm-like micellar gel. The non-Gaussian probability distribution functions (pdf) show a negative skewness very similar to the case of polymer solutions \cite{VSPRL09}. The Power Spectral Density (PSD) of $p(t)$ shows a power law behaviour with a decay exponent $\beta$ = -4 ($<-3$), a signature of elastic turbulence \cite{Fouxon}. But the decay exponent $\beta$ is very different ($\sim -2$) for constant shear-rate run which is not understood. 
We showed a universal scaling of the non-Gaussian pdf of $p(t)$ in terms of Generalized Gumbel (GG) distribution, observed in many cases of spatially averaged global measures for a diverse class of highly correlated systems \cite{Joubaud, electro1, electro2, 1/f, tokamak, thinfilm, Brey, glass,imbibition}. In-situ polarized light scattering studies clearly show a presence of spatially long ranged correlated dynamics in the turbulent regime although the temporal correlations are short ranged. We believe that the presence of this long ranged spatial correlation is responsible for the GG distribution of the spatially averaged global injected power. This kind of universality is similar to that observed for inertial turbulence \cite{BramwellNature, Pinton, Portelli}. The intermittent nature of the injected power below the mean value as compared to its value above the mean has been postulated to be associated with the nature of the elastic stress field in the system \cite{Burghelea}. We believe that our work will motivate further experimental and theoretical studies on the fascinating phenomenon of elastic turbulence in soft matter. 

AKS thanks CSIR for research support through the Bhatnagar Fellowship and SM thanks UGC for the Senior Research Fellowship.

\end{document}